\documentclass[aps,prl,twocolumn,superscriptaddress,nofootinbib]{revtex4-1}
\pdfoutput=1
\usepackage{amsmath,amssymb}
\usepackage{graphicx}
\usepackage{xcolor}
\usepackage[colorlinks=true,linkcolor=blue!60!black,citecolor=blue!60!black,urlcolor=blue!60!black]{hyperref}

\newcommand{\tM}{\widetilde{\mathcal{M}}}

\begin{document}

\title{Stabilizer R\'enyi Entropy Without a Phase Reference}

\author{Jinwei Chu}
\email{jinweichu@uchicago.edu}
\affiliation{Department of Physics, University of Chicago, Chicago, IL 60637, USA}
\affiliation{Departament de Física Quàntica i Astrofísica and Institut de Ciències del Cosmos (ICC), Universitat de Barcelona, Martí i Franquès 1, ES-08028, Barcelona, Spain}
\author{Savan Kharel}
\email{skharel@uchicago.edu}
\affiliation{Department of Physics, University of Chicago, Chicago, IL 60637, USA}

\date{\today}

\begin{abstract}
Stabilizer R\'enyi entropy quantifies magic relative to a complete Pauli frame. In settings such as helicity states of massless particles, unsynchronized qubit registers, energy eigenbases, and systems under superselection rules, the computational basis may be fixed while the transverse Pauli axes remain unspecified. The stabilizer R\'enyi entropy then depends on the choice of those axes. We define a phase-averaged stabilizer R\'enyi entropy by averaging the Pauli moments before taking the logarithm. It is exactly invariant under independent local rephasings. For pure states, it vanishes precisely when the state is a stabilizer state in every phase convention. For a broad class of sparse supports, it reduces to a closed-form function of the basis-state probabilities alone. Within this class, the ceiling for $n$ channels is $\ln[n^{3}/(6n-5)]$ for $n\geq3$ and $\ln\frac{7}{5}$ for two channels. We apply the invariant to tree-level gluon scattering at four and five points, where the helicity-basis phases are unspecified. It saturates the two- and three-channel ceilings: $\ln\frac{7}{5}$ at four points and $\ln\frac{27}{13}$ in $3\to2$ scattering. In same-helicity $2\to3$ scattering, it lies between a floor of $\ln\frac{8}{7}$ and a ceiling of $\ln\frac{184}{125}$, reached in the resolved soft limit and particular collinear limits, respectively. In every configuration studied, the invariant is independent of the fixed external colors.
\end{abstract}

\maketitle
\section*{Introduction}
Magic distinguishes nonstabilizer resources from states prepared by
Clifford  circuits acting on stabilizer inputs~\cite{Bravyi:2004isx,Leone:2021rzd}.
Its measurement on quantum processors~\cite{Oliviero:2022npjqi,Niroula:2023meg}
and investigation in scattering~\cite{White:2024nuc,Gargalionis:2025iqs,Chu:2026yxm}
make the specification of the Pauli frame a practical question. Nearly every such measurement, however, rests on an assumption: each
qubit comes with a full Pauli frame. In many physical settings, we fix the $\sigma_z$ eigenstates $\{|+\rangle,|-\rangle\}$ of each
qubit (e.g. helicity states of a massless particle, energy levels of an atom,
photon-number states of a mode) while leaving the transverse axes unspecified. 

Hence the basis is often defined only up to a phase for each qubit. Standard magic measures, e.g.,~the stabilizer R\'enyi entropies (SRE), are generically not invariant under the relative rephasings
$|+\rangle_i\to e^{i\theta_i}|+\rangle_i$ (with $|-\rangle_i$ fixed).\footnote{Locally minimized measures, such as nonlocal magic, are invariant under these rephasings~\cite{Cao:2023mzo,Cao:2024nrx,Qian:2025oit,Gargalionis:2026onv,
Robin:2025ymq}.}
For a Bell state $(|{+}{-}\rangle+|{-}{+}\rangle)/\sqrt2$ in one such basis, a direct
computation~\cite{Chu:2026yxm} gives the second SRE
\begin{equation}
\label{eq:bell}
M_2\;=\;\ln\frac{16}{14+2\cos 4(\theta_1-\theta_2)},
\end{equation}
which sweeps continuously from $0$ to $\ln\frac43$ as the relative phases $\theta_1$ and $\theta_2$ vary.

In this Letter we study an extension of the SRE adapted to settings with an unresolved phase frame and discuss its properties: it is exactly invariant. It vanishes if and only if the
state is a stabilizer state in \emph{every} phase convention. Also, for a
large family of states it reduces to a closed form in the basis-state
probabilities alone. Lastly, it captures the magic in typical phase frames. 

We then apply our invariant SRE to gluon scattering, where basis phases are often unspecified. While the methods are general, proposals to study magic at colliders~\cite{White:2024nuc} partly motivate this application. It is worth mentioning that gluon amplitudes, given their simplicity, have helped expose hidden mathematical symmetries and structures in quantum field theory~\cite{Parke:1986gb,Mangano:1990by}. In particular, we study four- and five-point gluon scattering, and find three surprising features: 1) saturated channel
ceilings, 2) a soft floor and a collinear ceiling, and 3) the exact cancellation of color-dependent factors. Further detailed derivations and calculations appear in
a companion paper~\cite{Chu:2026yxm}.
\section{A novel phase-invariant magic}
For a pure state $|\psi\rangle$ of $k$ qubits, the
$q$th SRE is
\begin{equation}
\label{Mq}
    M_q=\frac{1}{1-q}\ln\sum_{P}\frac{\langle\psi|P|\psi\rangle^{2q}}{2^{k}}\ ,
\end{equation} 
where the sum runs over
the $4^k$ Pauli strings~\cite{Leone:2021rzd}; it is computable from Pauli
moments with no optimization, has an operational
interpretation~\cite{Bittel:2025yhq}, and for pure states, is a magic
monotone for R\'enyi index $q\geq2$~\cite{Haug:2023hcs,Leone:2024lfr}.

In this Letter, we define a novel \emph{phase-invariant} SRE by averaging the Pauli moments over
the $k$ unresolved relative phases $\theta_i$, inside the logarithm:
\begin{equation}
\label{eq:inv}
\tM_q(|\psi\rangle)=\frac{1}{1-q}
\ln\!\left[\,\prod_{i=1}^{k}\int_{0}^{2\pi}\!\frac{d\theta_i}{2\pi}
\sum_{P}
\frac{\langle\psi_{\{\theta_i\}}|P|\psi_{\{\theta_i\}}\rangle^{2q}}{2^{k}}\right],
\end{equation}
where $|\psi_{\{\theta_i\}}\rangle$ denotes the same state with the basis rephased.

We will focus on the second SRE, $\tM_2$. The End Matter establishes the following properties:
\begin{enumerate}
\renewcommand{\labelenumi}{(\arabic{enumi})}

\item \emph{Invariance:} $\tM_2$ is exactly invariant under
all $U(1)^k$ rephasings, 

\item \emph{Vanishing criterion:} It vanishes if and only if
$|\psi\rangle$ is a stabilizer state in every phase convention.

\item \emph{Closed form from probabilities:}~Whenever the computational-basis components
of $|\psi\rangle=\sum_{i=1}^n c_i|s_i\rangle$ have pairwise-distinct flip
patterns--namely, for every pair $i < j$ with $c_i, c_j \neq 0$, the flip pattern $s_i \oplus s_j$ (where $\oplus$ denotes bitwise XOR) is distinct (e.g. $|\psi\rangle\propto |000\rangle+|101\rangle+|011\rangle$)---the torus average Eq.~\eqref{eq:inv} evaluates exactly to
\begin{equation}
\label{eq:master}
\tM_2\;=\;-\ln\!\left[\,6\Bigl(\sum_{i=1}^n |c_i|^4\Bigr)^{\!2}
-5\sum_{i=1}^n  |c_i|^8\right],
\end{equation}
a function of the probabilities $|c_i|^2$ alone. The unspecified phases have been integrated out, leaving an expression built entirely from populations,
which are measurable in a \emph{single} basis for a pure state of known support. 

Eq.~\eqref{eq:master} also leads to the   \emph{channel ceilings} (cf.\ the
maximal-magic question for the bare SRE Eq.~(\ref{Mq})~\cite{Liu:2025frx}): at equal
probabilities $|c_1|^2=|c_2|^2=\cdots=|c_n|^2=1/n$,
$\tM_2=\ln[n^3/(6n-5)]$---$\ln\frac87,\ln\frac{27}{13},\ln\frac{64}{19},
\ln 5,\dots$---and for $n\geq3$ this is the global maximum over the entire
family (proved in the End Matter), while for $n=2$ the maximum is instead
$\ln\frac75$, at populations $\tfrac12(1\pm\sqrt{3/7})$, with the equal-probability
point a local minimum.
\begin{figure*}[t]
\includegraphics[width=0.98\textwidth]{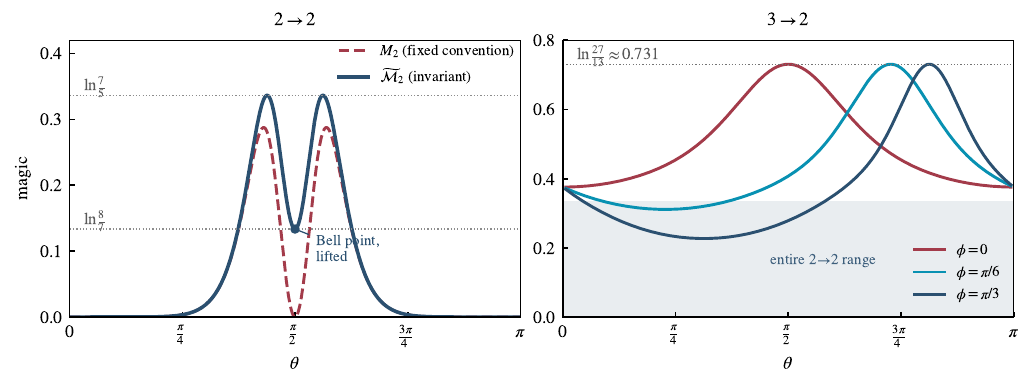}
\caption{\label{fig:one}%
The invariant at work in gluon scattering. \emph{Left:} $2\to2$. The bare
SRE $M_2$ in a fixed phase convention (dashed) vanishes at $\theta=\pi/2$,
where the final state is a Bell state~\cite{Gargalionis:2025iqs}; the invariant $\tM_2$ (solid) lifts it to $\ln\frac87$. \emph{Right:} $3\to2$
with initial helicities $|{+}{+}{-}\rangle$ and symmetric incoming momenta
(all-outgoing labels), at fixed azimuth $\phi$ (kinematics as in
Ref.~\cite{Chu:2026yxm}). The invariant never vanishes, and at every
$\phi$ reaches $\ln\frac{27}{13}\approx0.731$, a peak towering over the
entire $2\to2$ range (shaded).}
\end{figure*}

\item \emph{Typical- versus best-case}: By Jensen's inequality $\tM_2$ bounds
the phase-averaged bare SRE from below. Furthermore, in the two-channel example computed in \cite{Chu:2026yxm} the
gap never exceeds $0.006$, so $\tM_2$ closely tracks the mean. We expect this close agreement to hold more generally.

Our formulation is complementary to non-local or minimized
magic~\cite{Cao:2023mzo,Cao:2024nrx,Qian:2025oit,Gargalionis:2026onv,
Robin:2025ymq}, which optimizes over the full local-unitary group and
captures best-case magic.\footnote{For ensemble-averaged magic of random states see~\cite{Iannotti:2025lkb,Turkeshi:2023lqu}; the average here is over the
$U(1)^k$ orbit of a \emph{fixed} state.} The construction parallels the resource theory of quantum reference frames~\cite{Bartlett:2007rmp}: lacking a shared phase reference, one naturally works with quantities invariant under the unresolved $U(1)^k$ action. Eq.~\eqref{eq:inv} achieves this by averaging the Pauli moments over that orbit before taking the logarithm.
\end{enumerate}
\section{Application: gluon scattering}

A natural application is gluon scattering, where the helicity-basis phases are often unspecified. A massless gluon's two helicity states furnish
a qubit. Fixing external momenta and colors, the $S$-matrix of an $m\to k$ scattering process thus prepares the $k$-qubit state $|\psi\rangle=\sum_{h}\mathcal{A}_{m+k}(h)\,|h\rangle$, up to a normalization factor. We use tree-level color-dressed $(m+k)$-point amplitudes ($\mathcal{A}_{m+k}$) in the DDM decomposition~\cite{DelDuca:1999iql,DelDuca:1999rs}, built from Parke--Taylor partial amplitudes~\cite{Parke:1986gb,Mangano:1990by}; all
helicity labels $h$ are all-outgoing. The helicity states $|h\rangle$ are defined only up to
phases, and therefore it is useful to average over the rephasings as in
Eq.~\eqref{eq:inv}. 

Every nonvanishing four- or five-gluon tree amplitude is maximally helicity violating (MHV) or anti-MHV. Interestingly, every
case considered below satisfies the distinct
flip-pattern condition of Eq.~\eqref{eq:master}.

\emph{Four-point scattering:} First, consider $2\to2$
scattering (as in~\cite{Gargalionis:2025iqs}) with incoming helicities $|{+}{-}\rangle$, the final state is
$C|{+}{-}\rangle+D|{-}{+}\rangle$ with
$|C|/|D|=\tan^4\tfrac\theta2$ (where $\theta$ denotes the scattering angle in the center-of-mass frame), and
Eq.~\eqref{eq:master} gives
\begin{equation}
\label{eq:2to2}
\tM_2(\theta)=\ln\frac{\bigl(1+\tan^{8}\tfrac{\theta}{2}\bigr)^{4}}
{1+12\tan^{16}\tfrac{\theta}{2}+\tan^{32}\tfrac{\theta}{2}}\,.
\end{equation}
It vanishes only at $\theta=0$ and $\theta=\pi$. At the Bell point $\theta=\pi/2$ (at which $|C|=|D|$ and thus there exists a basis phase such that the state describes a Bell pair), the invariant is $\ln\frac87$ (Fig.~\ref{fig:one}, left). On the other hand, at $\theta\approx1.38$ and its mirror
$\pi-\theta$, the populations pass through $\tfrac12(1\pm\sqrt{3/7})$ and
the process \emph{saturates} the two-channel ceiling $\ln\frac75$. Thus four-point gluon scattering reaches the largest invariant allowed by two channels.
\\

\emph{Five-point scattering:} We want to note three interesting features exhibited by five-point scattering~\cite{Chu:2026yxm}:

\begin{enumerate}
\renewcommand{\labelenumi}{(\arabic{enumi})}
\item \emph{Three-channel saturation:} $3\to2$ scattering with incoming helicities $|{+}{+}{-}\rangle$ opens an anti-MHV channel, resulting in a final state of the form
\begin{equation}
C|{+}{-}\rangle+D|{-}{+}\rangle+E|{-}{-}\rangle.
\end{equation}
The invariant is independent of the fixed color configuration (see (3) below) and never vanishes in the configurations studied (Fig.~\ref{fig:one}, right). When the three probabilities coincide, it attains $\ln\frac{27}{13}\approx0.731$, saturating the three-channel ceiling. At this point, the bare SRE $M_2$ ranges over
$[\ln\frac95,\ln\frac94]$ as the phase convention varies~\cite{Chu:2026yxm}, with the maximum $\ln\frac94$ within two percent of the conjectured two-qubit maximum
$\ln\frac{16}{7}$~\cite{Liu:2025frx}.

\item \emph{A soft floor and a collinear ceiling:} In $2\to3$ scattering with incoming $|{+}{+}\rangle$, the outgoing state is
\begin{equation}
\begin{aligned}
B|{+}{-}{-}\rangle+C|{-}{+}{-}\rangle+D|{-}{-}{+}\rangle+E|{-}{-}{-}\rangle.
\end{aligned}
\end{equation}
In the center-of-mass (CM) frame, define the Dalitz energy fractions
$x_i=2E_i/E_{\rm CM}\in[0,1]$ for outgoing legs $i=3,4,5$, with $x_3+x_4+x_5=2$. The amplitude moduli obey
\begin{equation}
\label{eq:2to3}
\begin{aligned}
&|B|:|C|:|D|:|E|\\
&\qquad=(1-x_3)^2:(1-x_4)^2:(1-x_5)^2:1.
\end{aligned}
\end{equation}
Throughout the physical region, the invariant lies in the band $\ln\frac{8}{7}\leq\tM_2\leq\ln\frac{184}{125}$. Interestingly, the lower endpoint is a \emph{soft floor}; the upper endpoint is a ceiling approached in particular collinear limits (Fig.~\ref{fig:two}).

In the soft limit $E_3\to0$, the energy fractions approach $(x_3,x_4,x_5)=(0,1,1)$, and Eq.~\eqref{eq:2to3} reduces to $1:0:0:1$. The state factorizes into an equal-weight superposition of the soft gluon's helicities and the basis state $|{-}{-}\rangle$ of the other two gluons.  Thus, soft radiation does not erase the phase-invariant SRE; in this limit, the residual invariant is carried entirely by the soft gluon. This is counterintuitive, since one might naively expect the soft gluon to be negligible and hence to carry no magic.
\begin{figure}[t]
\includegraphics[width=0.98\columnwidth]{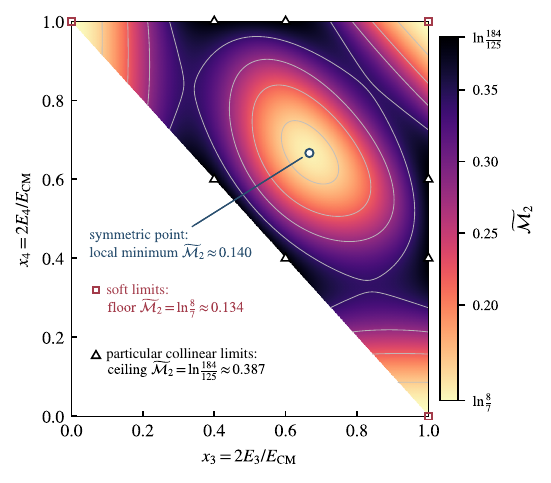}
\caption{\label{fig:two}%
The soft floor and collinear ceiling of three-qubit gluon magic from $2\to 3$ scattering: the
invariant over the Dalitz plane of outgoing energy fractions, from
Eqs.~\eqref{eq:master} and \eqref{eq:2to3}. At the vertices one gluon is
soft: the state degenerates to a single equatorial qubit carried by that
gluon, and the invariant approaches its floor $\ln\frac87$. The supremum
$\ln\frac{184}{125}$ is approached on the collinear boundary. The symmetric
point $x_3=x_4=x_5=\frac23$ is a local minimum,
$\ln\frac{4148928}{3606913}\approx0.140$.}
\end{figure}

On the collinear boundary $x_5=1$, the ceiling is approached at $(x_3,x_4)=(x_*,1-x_*)$ or $(1-x_*,x_*)$, where $x_*\approx 0.3976$. These two energy fractions are the roots in $(0,1)$ of
\[
\begin{aligned}
3x^6-9x^5-5x^4+25x^3-5x^2-9x+3=0.
\end{aligned}
\]
For this four-channel support, the channel ceiling is $\ln\frac{64}{19}\approx1.214$, attained at equal probabilities [Eq.~\eqref{eq:master}]. Equation~\eqref{eq:2to3} would then require $x_3=x_4=x_5=0$, forbidden by energy conservation. The Dalitz constraint thus keeps this same-helicity $2\to3$ process below a third of its channel ceiling, whereas $3\to2$ reaches its own.

\item \emph{Color independence:} Remarkably, for every configuration we consider, all color dependence cancels exactly. The MHV and anti-MHV channels involve conjugate color-kinematic sums, which become identical upon taking the absolute value, Eq.~(\ref{eq:master}). Consequently, the color dependence drops out entirely after normalization, and the magic of the gluon final states is determined solely by kinematics. The bare SRE, by contrast, retains color-dependent phases; only the phase-invariant measure is color-blind.
\end{enumerate}

\section*{Outlook}

A description of magic is incomplete until its reference frame is specified.
The phase-invariant SRE provides a diagnostic when the computational basis is
fixed but the transverse Pauli axes are unspecified. Interestingly, for the supports studied
here, it is determined entirely by populations. The scattering examples then
connect bounds on quantum states to the restrictions imposed by a physical
interaction: some processes attain their channel ceilings, while others remain
separated from them by kinematics.

The  next step is to understand which amplitude structures control this
distinction and the color cancellation found here. We need to consider higher multiplicities to test the persistence of color independence. On the
resource side, the task is to identify which operations compatible with the
retained frame make $\tM_q$ monotonic, and how its phase average relates to
measures optimized over local frames~\cite{Cao:2023mzo,Qian:2025oit}.

Collider applications must connect this conditional pure-state diagnostic to
measurements that average over color and are affected by hadronization.
Recent studies of spin correlations and quantum information at colliders
motivate that task~\cite{Afik:2020onf,ATLAS:2023fsd,CMS:2024pts,Cheng:2025zaw,Fucilla:2026mkg}.
For quantum registers, Eq.~\eqref{eq:time} suggests a route through
time-averaged Pauli moments. Establishing its accuracy at finite times and
its sensitivity to noise would turn the ideal identity into a measurement
protocol. Such a protocol would allow magic to be studied with the phase
information that an experiment actually supplies.

\subsection{Acknowledgments}
We thank Soner Albayrak, Lukas Graf, Qiyuan Hu, Sam Li, and Dongheng Qian for valuable discussions and feedback. JC thanks Westlake University and B(eijing)IMSA for hospitality during the conclusion of this
work. SK gratefully acknowledges CERN for its invitation
and hospitality, and in particular the TH Institute on Quantum Simulation and
Computation in High-Energy Physics, where this project originated.

\bibliography{paper}

\vspace{2mm}
\subsection{End Matter}
\subsubsection*{Invariance, vanishing, and the Jensen bound}
Write $\Pi(\{\theta_i\})=\frac{1}{2^k}\sum_P
\langle\psi_{\{\theta_i\}}|P|\psi_{\{\theta_i\}}\rangle^4$
for the stabilizer purity of the rephased state. Then
$\tM_2=-\ln\langle\Pi\rangle_{\{\theta_i\}}$, where
$\langle\cdot\rangle_{\{\theta_i\}}$ is the normalized Haar average over $U(1)^k$.

\emph{Invariance:} Rephasing by $\{\alpha_i\}$ translates
$\theta_i\to\theta_i+\alpha_i$ under the integral. Translation invariance
of the Haar measure leaves the average unchanged.

\emph{Vanishing criterion:} For a pure state, $\Pi(\{\theta_i\})\leq1$, with
equality precisely for stabilizer states~\cite{Leone:2021rzd}. Thus
$\langle\Pi\rangle_{\{\theta_i\}}=1$ requires $\Pi(\{\theta_i\})=1$ almost everywhere.
Continuity extends this equality to every phase convention. Hence
$\tM_2=0$ when the state is a stabilizer state in every convention.

\emph{Jensen bound:} Convexity of $-\ln$ gives
\begin{equation}
\tM_2=-\ln\langle\Pi\rangle_{\{\theta_i\}}
\leq\langle-\ln\Pi\rangle_{\{\theta_i\}}.
\end{equation}
The right-hand side is the phase-averaged bare SRE. For the two-channel
\begin{equation}
\Pi(\Delta\theta)=
\frac{1+[12+2\cos(4\Delta\theta)]r^4+r^8}{(1+r^2)^4},
\end{equation}
where $r$ is the ratio of amplitude moduli and $\Delta\theta=\theta_1-\theta_2$ is their
relative phase. The gap between the two averages is below $0.006$ for
this family~\cite{Chu:2026yxm}.

\subsubsection*{Closed form on sparse supports}
For $|\psi\rangle=\sum_i c_i|s_i\rangle$, expand each Pauli moment and
average over the torus. The distinct-flip condition requires different
unordered pairs $i<j$ to have different bitwise differences $s_i\oplus s_j$.
Equivalently, no four distinct support strings have vanishing bitwise XOR.
Under this condition, the phase-sensitive terms integrate to zero, and the
surviving terms combine into Eq.~\eqref{eq:master}. When the condition fails,
additional terms can survive, so the probability-only formula need not apply.

For two channels with populations $(r^2,1)/(1+r^2)$, the result reduces to
\begin{equation}
\tM_2=\ln\frac{(1+r^2)^4}{1+12r^4+r^8}.
\end{equation}
The $2\to2$ curve in Fig.~\ref{fig:one} follows by setting
$r=\tan^4\tfrac\theta2$.

\subsubsection*{Channel ceilings}
Set $p_i=|c_i|^2$. Maximizing Eq.~\eqref{eq:master} amounts to minimizing
\begin{equation}
F(p)=6\Bigl(\sum_i p_i^2\Bigr)^2-5\sum_i p_i^4,
\qquad \sum_i p_i=1.
\end{equation}
At equal populations, $F=(6n-5)/n^3$. Interior stationary points obey
\begin{equation}
\label{stat}
    (p_i-p_j)\bigl[6\!\sum_s p_s^2-5(p_i^2+p_ip_j+p_j^2)\bigr]=0
\end{equation}
for every pair $i,j$. If $p_i\neq p_j$ and $p_i\neq p_k$, subtracting
the corresponding equations gives $(p_j-p_k)(p_i+p_j+p_k)=0$.
The positive populations take at most two distinct values.

Suppose $m$ populations equal $a$ and the remaining $n-m$ equal $b\neq a$.
Equation~\eqref{stat} then requires $(6m-5)a^2-5ab+[6(n-m)-5]b^2=0$.
The discriminant satisfies
\begin{equation}
\begin{aligned}
\Delta&=25-4(6m-5)[6(n-m)-5]\\
&\leq69-24n<0\qquad(n\geq3).
\end{aligned}
\end{equation}
For $n\geq3$, equal populations therefore give the only interior stationary
point. For $n=2$, writing $u=p_1p_2$ gives
\begin{equation}
F=1-4u+14u^2=\frac57+14\left(u-\frac17\right)^2,
\end{equation}
so the minimum is $5/7$, at $p_1p_2=1/7$.

To summarize, the global ceilings are
\begin{equation}
\tM_2^{\max}=
\begin{cases}
\ln(7/5), & n=2,\\
\ln[n^3/(6n-5)], & n\geq3.
\end{cases}
\end{equation}
Note that any three distinct basis strings satisfy the distinct-flip condition, so
$\ln(27/13)$ bounds every three-channel state.

\subsubsection*{Dynamical analog}
For qubits evolving under local fields,
\begin{equation}
H=\sum_{i=1}^k\frac{\omega_i}{2}\,\sigma_z^{(i)},
\end{equation}
free evolution sweeps out the local phase orbit. If the frequencies are
rationally independent, Weyl's theorem equates the long-time average with
the Haar average over $U(1)^k$:
\begin{equation}
\label{eq:time}
\lim_{T\to\infty}\frac{1}{T}\!\int_0^T\!\!dt\sum_{P}
\frac{\langle\psi(t)|P|\psi(t)\rangle^{4}}{2^{k}}
=e^{-\tM_2(|\psi\rangle)}.
\end{equation}
Rational independence is sufficient. More generally, the equality holds
when the stabilizer purity contains no
resonance among the frequencies. The invariant is thus analogous to minus the logarithm of the time-averaged
stabilizer purity of a freely precessing register.

\end{document}